\documentclass[conference]{IEEEtran}
\IEEEoverridecommandlockouts
\usepackage{amsmath,amssymb,amsfonts}
\usepackage[table,xcdraw]{xcolor}
\usepackage{progressbar}
\usepackage{tikz}
\usepackage{adjustbox}
\usepackage{tcolorbox}
\usepackage{fontawesome}
\usepackage{multirow}
\usepackage[normalem]{ulem}
\usepackage{tabularray}
\definecolor{Gallery}{rgb}{0.937,0.937,0.937}
\usepackage[numbers]{natbib}

\usepackage{algorithmic}
\usepackage{graphicx}
\usepackage{textcomp}
\usepackage{enumitem}
\usepackage{url}
\usepackage{array}
\usepackage{xspace}
\usepackage{soul}
\usepackage{xcolor}
\definecolor{softyellow}{HTML}{fceabb}
\definecolor{softorange}{HTML}{f8c291}
\definecolor{softred}{HTML}{f19066}
\definecolor{darkred}{HTML}{d56c6c}
\usepackage{tabularx}
\usepackage{booktabs}
\usepackage{makecell}
\usepackage{array}
\usepackage{multirow}
\usepackage{pifont}
\usepackage{subcaption}
\usepackage{booktabs}
\usepackage{soul}

\usepackage[bottom]{footmisc}
\newcolumntype{Y}{>{\raggedright\arraybackslash}X}

\newif\ifuseMicrosoft
\useMicrosoftfalse  

\ifuseMicrosoft
  \newcommand{\companyname}{Microsoft}
\else
  \newcommand{\companyname}{Microsoft}
\fi

\newif\ifdefineDISD
\defineDISDfalse

\ifdefineDISD
  \newcommand{\DISD}{\textbf{\textit{Cross-Disciplinary Software Development (CDSD)}}\xspace}
\else
  \newcommand{\DISD}{CDSD\xspace}
\fi

\usetikzlibrary{fit} 

\newcommand{\boldification}[1]{\ifdraft\indent ** \textbf{#1}** \\ \indent\else\relax\fi}
\newif\ifdraft
\draftfalse 

\newif\ifdraft
\draftfalse 

\def\BibTeX{{\rm B\kern-.05em{\sc i\kern-.025em b}\kern-.08em
    T\kern-.1667em\lower.7ex\hbox{E}\kern-.125emX}}

\tcbset{
    mybox/.style={
        colback=gray!10, 
        colframe=black, 
        width=\linewidth, 
        boxrule=1mm, 
        leftrule=1mm, 
        rightrule=1mm, 
        toprule=0mm, 
        bottomrule=0mm, 
        sharp corners, 
        boxsep=0pt, 
        left=5pt, 
        right=9pt, 
        top=5pt, 
        bottom=5pt 
    }
}

\newcommand{\four}[4]{%
  \begin{tikzpicture}[baseline]
    \pgfmathsetmacro{\barwidth}{3.5} 
    \pgfmathsetmacro{\widthSD}{#1*\barwidth/100}
    \pgfmathsetmacro{\widthD}{#2*\barwidth/100}
    \pgfmathsetmacro{\widthA}{#3*\barwidth/100}
    \pgfmathsetmacro{\widthSA}{#4*\barwidth/100}

    \fill[softyellow] (0,0) rectangle (\widthSD,0.5); 
    \fill[softorange] (\widthSD,0) rectangle (\widthSD+\widthD,0.5); 
    \fill[softred] (\widthSD+\widthD,0) rectangle (\widthSD+\widthD+\widthA,0.5); 
    \fill[darkred] (\widthSD+\widthD+\widthA,0) rectangle (\widthSD+\widthD+\widthA+\widthSA,0.5); 

    \draw[black] (0,0) rectangle (\barwidth,0.5);

    \pgfmathparse{#1} \ifdim\pgfmathresult pt > 15pt
      \node[font=\scriptsize] at (\widthSD/2, 0.25) {#1\%};
    \fi
    \pgfmathparse{#2} \ifdim\pgfmathresult pt > 15pt
      \node[font=\scriptsize] at (\widthSD+\widthD/2, 0.25) {#2\%};
    \fi
    \pgfmathparse{#3} \ifdim\pgfmathresult pt > 15pt
      \node[font=\scriptsize] at (\widthSD+\widthD+\widthA/2, 0.25) {#3\%};
    \fi
    \pgfmathparse{#4} \ifdim\pgfmathresult pt > 15pt
      \node[font=\scriptsize, text=white] at (\widthSD+\widthD+\widthA+\widthSA/2, 0.25) {#4\%};
    \fi
  \end{tikzpicture}%
}

\definecolor{StronglyDisagree}{HTML}{e78c8c} 
\definecolor{Disagree}{HTML}{f5c5a0} 
\definecolor{Neutral}{HTML}{ffe08c} 
\definecolor{Agree}{HTML}{b1dbaa} 
\definecolor{StronglyAgree}{HTML}{77cfa6} 

\newcommand{\five}[5]{%
  \makebox[4cm][c]{
    \begin{tikzpicture} 
      \pgfmathsetmacro{\barwidth}{3} 
      \pgfmathsetmacro{\widthSD}{#1*\barwidth/100}
      \pgfmathsetmacro{\widthD}{#2*\barwidth/100}
      \pgfmathsetmacro{\widthN}{#3*\barwidth/100}
      \pgfmathsetmacro{\widthA}{#4*\barwidth/100}
      \pgfmathsetmacro{\widthSA}{#5*\barwidth/100}

      \fill[StronglyDisagree] (0,0) rectangle (\widthSD,0.5); 
      \fill[Disagree] (\widthSD,0) rectangle (\widthSD+\widthD,0.5); 
      \fill[Neutral] (\widthSD+\widthD,0) rectangle (\widthSD+\widthD+\widthN,0.5); 
      \fill[Agree] (\widthSD+\widthD+\widthN,0) rectangle (\widthSD+\widthD+\widthN+\widthA,0.5); 
      \fill[StronglyAgree] (\widthSD+\widthD+\widthN+\widthA,0) rectangle (\widthSD+\widthD+\widthN+\widthA+\widthSA,0.5); 

      \draw[black] (0,0) rectangle (\barwidth,0.5);

      \pgfmathparse{#1} \ifdim\pgfmathresult pt > 15pt
        \node[font=\scriptsize] at (\widthSD/2, 0.25) {#1\%};
      \fi
      \pgfmathparse{#2} \ifdim\pgfmathresult pt > 15pt
        \node[font=\scriptsize] at (\widthSD+\widthD/2, 0.25) {#2\%};
      \fi
      \pgfmathparse{#3} \ifdim\pgfmathresult pt > 15pt
        \node[font=\scriptsize] at (\widthSD+\widthD+\widthN/2, 0.25) {#3\%};
      \fi
      \pgfmathparse{#4} \ifdim\pgfmathresult pt > 15pt
        \node[font=\scriptsize] at (\widthSD+\widthD+\widthN+\widthA/2, 0.25) {#4\%};
      \fi
      \pgfmathparse{#5} \ifdim\pgfmathresult pt > 15pt
        \node[font=\scriptsize] at (\widthSD+\widthD+\widthN+\widthA+\widthSA/2, 0.25) {#5\%};
      \fi
    \end{tikzpicture}%
  }%
}

\begin{document}

\title{When Domains Collide: An Activity Theory Exploration of Cross-Disciplinary Collaboration \thanks{979-8-3315-9147-2/25/\$31.00 ©2025 IEEE}} 

\author{\small%
    Zixuan Feng\textsuperscript{1}, 
    Thomas Zimmermann\textsuperscript{2}, 
    Lorenzo Pisani\textsuperscript{3}, 
    Christopher Gooley\textsuperscript{3}, 
    Jeremiah Wander\textsuperscript{3}, 
    Anita Sarma\textsuperscript{1}\\
    \textsuperscript{1}Oregon State University, USA, \{fengzi, Anita.Sarma\}@oregonstate.edu;\\
    \textsuperscript{2}University of California, Irvine, USA, tzimmer@uci.edu;\\
    \textsuperscript{3}Microsoft Research, USA, \{lorenzo.pisani, gooley, miah\}@microsoft.com;\\
}

\maketitle

\begin{abstract}
\textit{Background:} Software development teams are increasingly diverse, embedded, and cross-disciplinary. Domain experts (DEs) from different disciplines collaborate with professional software developers (SDEs), bringing complementary expertise in creating and maintaining complex production software. However, contested expectations, divergent problem-solving perspectives, and conflicting priorities lead to friction.  
\textit{Aims:} This study aims to investigate the dynamics of emerging collaboration of cross-disciplinary software development (\DISD) by exploring the expectations held by DEs and SDEs and understanding how these frictions manifest in practice. \textit{Method:} We utilize Activity Theory (AT), a well-established socio-technical framework, as an analytical lens in a grounded, empirical investigation, conducted through a mixed-method study involving 24 interviews (12 DEs and 12 SDEs) and a large-scale validation survey with 293 participants (161 DEs and 132 SDEs). \textit{Results:} We conceptualize and empirically ground the CDSD dynamics. We identified eight expectations held by SDEs and six by DEs. By mapping these expectations to AT components, we revealed 21 frictions in CDSD and illustrated where and how they arise.  \textit{Conclusions:} This study offers a theoretical lens for understanding the dynamics and frictions in CDSD and provides actionable insights for future research, practitioners, and infrastructure design.

\end{abstract}

\begin{IEEEkeywords}
Cross-disciplinary Collaboration, Activity Theory, Mixed-Methods
\end{IEEEkeywords}
\section{Introduction}
Consider a team of three software engineers and three medical physicists co-developing an MRI image processing software for early cancer detection with fluid roles; physicists configure the CI/CD pipeline, while software engineers implement the image processing models, with all team members sharing code review and deployment responsibilities. Such domain expert embedded, ``Cross-Disciplinary Software Development (\DISD)'' teams,  where domain experts and software engineers co-own the product and share responsibilities, promise richer expertise and faster innovation. Yet subtle differences in priorities and tooling literacy can produce misaligned expectations, which then manifest as friction in collaboration.

While \DISD teams are becoming increasingly common—ranging from biology \cite{list2017ten, snir2021democratizing} to health \cite{siddique2024leveraging, weber2013software} to the automotive industry \cite{dikmen2017trust}, among others, our understanding of how they operate is still unclear. Prior works on cross‐disciplinary work have focused on Domain Experts (DE) serving as clients \cite{baxter2022collaborative, li2017cross}, in peripheral roles (e.g., data scientists as isolated consultants) \cite{kim2016emerging, kim2017data}, or participating in hand‐off collaboration models \cite{nahar2022collaboration} (e.g., siloed DEs produce Machine Learning models that are integrated by Software Developers (SDE)), leaving the dynamics of fully embedded, co‐owned cross-disciplinary collaboration largely unexplored.

To unpack the complex, multi-actor, role-fluid interactions and frictions in \DISD, we use the lens of Activity Theory (AT) \cite{engestrom1999activity, de2003using}, a well-established socio-technical framework that conceptualizes work as an interplay among subjects (SDEs, DEs), tools (infrastructure, pipelines), rules (engineering practices), community, division of labor, and shared objects \cite{de2003using} to produce the outcome of project success. AT is especially suited for our investigation as it explicitly models mediating artifacts, rules, and division of labor, enabling it to reveal how mismatches in expectations and priorities cause frictions. 
We use the lens of AT to investigate:
%
\textbf{RQ1}:  \emph{How do SDEs and DEs coordinate through tools, rules, and shared artifacts when co-developing software in \DISD teams?}
\textbf{RQ2}: \emph{What frictions arise from mismatched expectations in \DISD collaborations?}

We answer these questions through a mixed-methods study \cite{storey2024guiding}, 24 interviews spanning 12 SDEs and 12 DEs from diverse organizations, projects, and regions within \companyname{}, followed by a large-scale survey validation of 294 Open Source Software developers with experience in \DISD. By grounding our mixed-methods study in AT, we identified 14 expectations each group holds for the other, and a set of 21 collaboration frictions arising from mismatched expectations.

The contribution of our work is threefold: (1) We apply AT to conceptualize and empirically ground the fluid team dynamics that emerge in cross-disciplinary software development, offering a theoretical lens to understand how shared ownership and blurred roles unfold in practice.

(2) We map expectations of DEs and SDEs to AT components to surface how and where frictions arise, revealing that collaboration breakdowns stem from systemic contradictions within the activity system, such as misaligned expectations in rules, tools, and the division of labor. Our findings equip SE researchers and practitioners with a diagnostic framework to anticipate, surface, and resolve contestation before it escalates.

(3) We leverage AT to identify collaboration friction hotspots in \DISD, particularly around rules, tools, and community to offer actionable insights to: (i) enable practitioners focus efforts on high-impact areas (e.g., setting up shared validation practices that balance SDE rigor with DE agility), (ii) guide infrastructure designers to develop adaptive tools (e.g., lightweight rule enforcement mechanisms), and (iii) inspire software engineering researchers to explore new coordination mechanisms (e.g., reimagining Agile to accommodate contributors with limited software engineering backgrounds).

Our findings suggest that collaboration in \DISD fails quietly, at the boundary of assumptions, not ability; \textit{``There’s a lot of philosophical differences in software development practices.''} [DE8] (interview participant). Our work makes those boundaries visible and offers a structured AT lens to navigate them.

To facilitate reproducibility, we include the interview guide, codebook, and survey questions in the supplemental material \cite{runeson2024conceptual}. We do not include interview transcripts or survey responses to maintain participant confidentiality and comply with our IRB protocol.

\label{sec:intro}

 \section{Background}
Collaboration in software development teams is a long-studied phenomenon \cite{brooks1995mythical, whitehead2007collaboration}, but as technology and practices evolve, so does collaboration and its associated frictions. We focus on the collaboration aspect of how SDE works with DEs, which can be across the team border as clients or, as in more recent situations, embedded within the team itself with a shared outcome and client focus. We categorized the different types of collaborations into four models.

\textbf{Professional Collaboration} typically involves software professionals \cite{brooks1995mythical, lehman1980programs}, often working for clients in industries such as finance, transportation, and government \cite{steinmueller1995us, royce1987managing}. Responsibility for creating the product and its ownership remains entirely with the software team. Collaboration to build the software is internal to the development team, with checkpoints with clients as required in the development process \cite{boehm1986spiral, abbott1981software}.

\textbf{Adjacent Collaboration} introduces DEs (e.g., data scientists) as part of the software development teams, but relegates them to peripheral roles \cite{kim2016emerging, kim2017data}. In the case of data scientists, they are primarily involved in data-related helper tasks, such as processing, analyzing, and modeling data to inform the system design \cite{kim2017data, davoudian2020big}, and their involvement remains isolated from the main software development workflows \cite{sculley2015hidden}.

\textbf{Interfaced Collaboration} is a more cross-disciplinary collaboration type \cite{ nahar2022collaboration}, where DE teams, such as machine learning engineers, work with SDE teams at specific stages~\cite{lwakatare2019taxonomy}. This form of collaboration involved more interactions between the DEs and SDEs, but their development roles were still siloed: DEs focused on building models, while SDEs were responsible for production and deployment  \cite{nahar2022collaboration}. Such phase-specific engagements creates fragmented ownership, misaligned expectations, and integration bottlenecks \cite{nahar2022collaboration}. Dependency management across the teams is uneven and poorly coordinated, which limits collaboration \cite{amershi2019software, nahar2022collaboration}.

\textbf{Embedded Collaboration }(\DISD). A more recent collaboration pattern facilitated by tools such as generative AI that allows DEs, such as aerospace engineers and healthcare professionals, to work with SDEs within the same team to co-develop software across the entire development lifecycle \cite{li2017cross}. In this model, dependency is bi-directional, ownership and responsibility are collectively held, with a unified goal and continuous integration of domain rigor and software practices.

\textbf{Collaboration Frictions.}
Despite technological advancements, no ``silver bullet'' \cite{brooks1995mythical} exists for resolving collaboration frictions. As collaboration models evolve and DEs' participation gets more integrated with that of SDEs, so do the collaboration frictions.  These frictions arise because of differing expertise and philosophies \cite{kim2017data, nahar2022collaboration}. \citet{murphy2014cowboys} investigated how game SDEs worked with SDEs to create gaming products and found that SDEs from different backgrounds had different testing philosophies.

These differences only amplify as the differences in backgrounds increase. \citet{kim2017data} found that challenges between SDEs and data scientists in \emph{adjacent collaborations} occurred with data scientists' responsibilities. First, about the data itself, including data quality and data pipeline. Second, about their role in the team. Data scientists often served as data providers with limited access, authority, and voice; their work was frequently misunderstood, treated unequally, and they needed to continually justify the value of data science contributions, leading to communication difficulties \cite{kim2017data}.

\citet{nahar2022collaboration} identified challenges in \emph{interfaced collaborations}, where problems occur during the integration handoff in between-team workflows. Model (Data) teams primarily focused on model development, but often had unclear responsibilities for deployment, whereas SDEs were responsible for product development and system integration. This division led to fragmented ownership, with the integration process appearing as a black box to both sides \cite{nahar2022collaboration}.

Past works have investigated the challenges that end-user programmers, DEs creating software for their \emph{own use}, face when adopting software engineering practices, such as testing, version management, debugging, and maintaining code quality \cite{barricelli2019end}. One recurring theme is that their goals are in creating the software to meet their own demand and not about the maintainability or reproducibility of the software \cite{ko2011state}.

This work investigates the \emph{embedded collaboration} model and its ensuing frictions. The dynamics of how DEs and SDEs collaborate are underexplored, including which frictions are prevalent and what factors create these frictions. With differing expertise and philosophies, work together, expectation mismatches are bound to arise \cite{nahar2022collaboration, kim2017data}; but, what are the expectations of each side \cite{zhang2019capabilities} and how mismatches in expectations create collaboration frictions is unknown.

\textbf{Activity Theory (AT) Lens to unpack \DISD.} 
Several theories have been used to analyze team collaboration, including theories such as Social Interdependence Theory \cite{rusbult2008we}, Collaboration Theory \cite{colbry2014collaboration}, and Social Exchange Theory \cite{cook2013social}.

We use Activity Theory (AT) as our lens to investigate embedded collaboration as manifested in \DISD. We do so since AT is a well-established framework initially developed by \citet{engestrom1999activity} that provides a lens for analyzing complex human activity through the interactions among subjects, objects, tools, rules, community, and division of labor. It has since been widely used to analyze collaboration and identify friction across domains \cite{zahedi2017understanding, mccance2023using, nardi1996studying}.

\begin{figure}[!htbp]
    \centering
    \includegraphics[width=0.7\columnwidth]{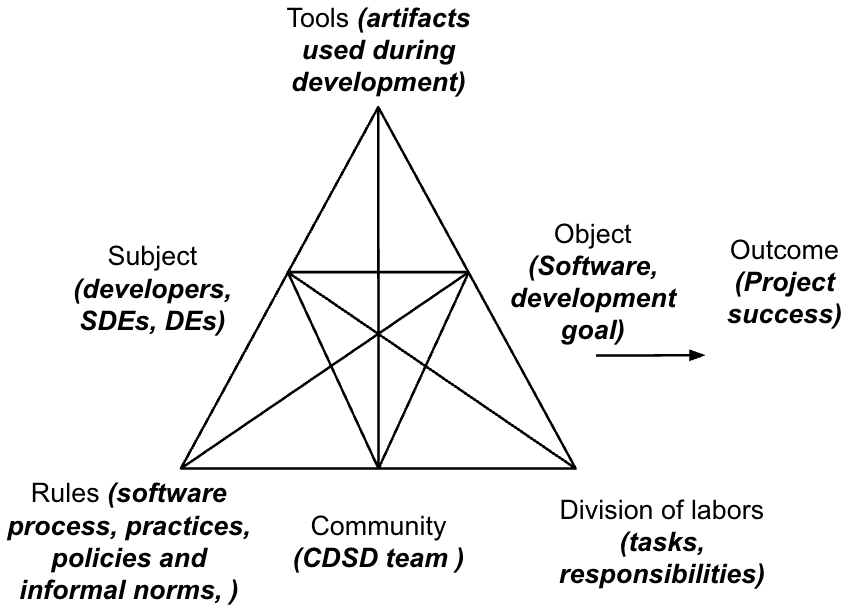}
    \caption{Using Activity to Understand \DISD Team}
    \label{fig:at}
    \vspace{-3mm}
\end{figure}

\boldification {We use the existing AT framework in software engineering.}
To apply AT in \DISD, we build on the established mapping from \citet{de2003using}, which applied second-generation AT \cite{engestrom1999activity} to analyze tensions and contradictions in conventional software development teams. 

In \DISD, as shown in Figure~\ref{fig:at}, SDEs and DEs work together as developers (\emph{subject}) toward building software systems (\emph{object}). The collaboration is driven by shared ownership and a common pursuit of the \emph{outcome}. The \emph{division of labor} represents how tasks are distributed across team members. The \emph{tools} mediating their work include various development artifacts (e.g., code repositories, configuration management). \emph{Rules} governing the collaboration encompass both formal practices (e.g., software engineering practices) and informal conventions (e.g., ad hoc testing, a ``move fast'' culture). The \emph{community} includes all developers engaged in the product development effort.

\label{sec:related}

\section{Study Design}

\boldification{Our study is guided by the AT and consists of 2 phases}
We used Activity Theory (AT) to guide our mixed-method \cite{storey2024guiding} study to systematically investigate the different activities and interactions among SDEs and DEs, as well as the expectations and frictions in \DISD to answer our research questions.

\textit{Study 1} was an exploratory interview conducted at \companyname{}. \companyname{} is an American multinational software company includes multiple organizations, each leading the development of different products ranging from cloud services and video games to research. Each organization has various projects, and employees can contribute to multiple projects. An emerging theme in these development teams is that DEs and SDEs are integrated with the shared responsibility to create and deploy products. For example, one team [DE7, SDE10] was developing a biomedical image processing system.

\textit{Study 2} was a confirmatory survey designed to strengthen the credibility and generalizability of the results identified in Study 1, following the mixed-methods principles of triangulation, development, and expansion \citep{storey2024guiding}. It included 9 participants from \companyname{} and 284 from OSS projects.

\subsection{Study 1: AT-Guided Exploratory Interview}

We used the existing AT framework in software engineering  \cite{de2003using} (as explained in Section \ref{sec:related}) to guide the design of our interviews. The interview was designed to be semi-structured, allowing the conversation to evolve organically, enabling us to understand the \DISD systematically \cite{srivastava2009practical}.

To understand SDEs and DEs experiences before making cross-group interpretations \cite{kramer1994sinister}, we adopted an incremental approach by interviewing SDEs and DEs separately rather than simultaneously. Specifically, we conducted interviews with one group first (i.e., SDEs), performing ongoing analysis until we observed convergence \cite{charmaz2006constructing}. We then switched to the other group (i.e., DEs) and followed a similar incremental process until convergence was again observed.

\textbf{SDE Interview Design.} was designed based on the components of AT. In our study, the \emph{subjects} are the SDEs themselves. We collected additional demographic (gender, region, and years of experience as SDEs) and their roles within their projects. We then asked questions related to the \emph{object}, such as the shared goal or software artifacts they build (e.g., \textit{``What is the primary goal or product of your current project?''}). Next, we asked questions related to the \emph{tools} they used, including programming languages, development platforms, tools, and collaboration infrastructure. For the \emph{rules} component, we asked about formal and informal norms, including adherence to agreed-upon software engineering practice, such as clean code or SOLID principles \cite{martin2009clean}.

Regarding \emph{community}, our questions focused on how SDEs and DEs interact, including collaboration dynamics, mutual expectations, perceptions of efficiency, frictions, productivity, and success of projects. For example, we asked what expectations they have for DEs and what frictions they experience.

As for the \emph{division of labor}, we asked participants to describe how tasks were distributed, with follow-up questions about how dependencies were managed. For the \emph{outcomes}, we included questions related to participants’ perceptions of project success and the quality of their work. To guide a more systematic understanding of outcomes, we used the ISO/IEC 25010 quality model \cite{ISO25010} to frame questions, such as functional suitability, performance efficiency, maintainability, and reliability. Refer to the supplementary materials for the details of interview scripts \cite{supply}.

\textbf{DE Interview Questions.} 
In addition to reusing the questions discussed above, we also asked questions based on the qualitative findings from the SDE interviews. For example, SDEs mentioned relaxed code quality contributed by DEs, we asked DEs about how they manage the code quality and whether there are any frictions regarding code quality.

\textbf{Pilot interviews.} 
Before conducting the main interviews, we piloted six interviews to collect feedback on the interview design within an organization at \companyname{}, involving participants with experience in \DISD{} (3 SDEs, 3 DEs). We collected feedback and refined the interview questions accordingly. For example, pilot participants suggested adding follow-up questions to understand the reasons behind expectations.

\begin{table}[!tbp]
\caption{PARTICIPANT INFORMATION}
\centering
\resizebox{\columnwidth}{!}{
\begin{tabular}{lllllll}
\multicolumn{3}{c}{\cellcolor[HTML]{EFEFEF}\textbf{Software Engineer}} &  & \multicolumn{3}{c}{\cellcolor[HTML]{EFEFEF}\textbf{Domain Experts}} \\
\textbf{ID} & \textbf{Gender} & \textbf{Title} &  & \textbf{ID} & \textbf{Gender} & \textbf{Title} \\
\cellcolor[HTML]{EFEFEF}SDE1 & \cellcolor[HTML]{EFEFEF}Man & \cellcolor[HTML]{EFEFEF}Pr. SDE &  & \cellcolor[HTML]{EFEFEF}DE1 & \cellcolor[HTML]{EFEFEF}Man & \cellcolor[HTML]{EFEFEF}Pr. Researcher in Physics \\
SDE2 & Man & Pr. SDE &  & DE2 & Man & Pr. Researcher in Electr. Eng \\
\cellcolor[HTML]{EFEFEF}SDE3 & \cellcolor[HTML]{EFEFEF}Man & \cellcolor[HTML]{EFEFEF}Pr. SDE &  & \cellcolor[HTML]{EFEFEF}DE3 & \cellcolor[HTML]{EFEFEF}Man & \cellcolor[HTML]{EFEFEF}Pr. Researcher in Biology \\
SDE4 & Man & Sr. SDE &  & DE4 & Man & Sr. Pr. Researcher in Machine Learning \\
\cellcolor[HTML]{EFEFEF}SDE5 & \cellcolor[HTML]{EFEFEF}Woman & \cellcolor[HTML]{EFEFEF}Pr. SDE &  & \cellcolor[HTML]{EFEFEF}DE5 & \cellcolor[HTML]{EFEFEF}Man & \cellcolor[HTML]{EFEFEF}Pr. Data Scientist \\
SDE6 & Woman & Pr. SDE &  & DE6 & Man & Sr. Scientist in Genomics \\
\cellcolor[HTML]{EFEFEF}SDE7 & \cellcolor[HTML]{EFEFEF}Woman & \cellcolor[HTML]{EFEFEF}Pr. SDE &  & \cellcolor[HTML]{EFEFEF}DE7 & \cellcolor[HTML]{EFEFEF}Man & \cellcolor[HTML]{EFEFEF}Sr. Director* in Bio-statistics \\
SDE8 & Woman & Sr. SDE &  & DE8 & Man & Sr. Applied Scientist in Bio-statistics\\
\cellcolor[HTML]{EFEFEF}SDE9 & \cellcolor[HTML]{EFEFEF}Man & \cellcolor[HTML]{EFEFEF}Pr. SDE &  & \cellcolor[HTML]{EFEFEF}DE9 & \cellcolor[HTML]{EFEFEF}Woman & \cellcolor[HTML]{EFEFEF}Sr. Researcher in Data Science \\
SDE10 & Man & SDE* &  & DE10 & Man & Pr. Researcher in Computational Imaging \\
\cellcolor[HTML]{EFEFEF}SDE11 & \cellcolor[HTML]{EFEFEF}Man & \cellcolor[HTML]{EFEFEF}Sr. SDE &  & \cellcolor[HTML]{EFEFEF}DE11 & \cellcolor[HTML]{EFEFEF}Woman & \cellcolor[HTML]{EFEFEF}Sr. Researcher in Machine Learning \\
SDE12 & Man & Sr. SDE &  & DE12 & Woman & Sr. Scientist in Biostatistics \\
\multicolumn{7}{l}{\cellcolor[HTML]{EFEFEF}SDE* is equivalent to Sr. SDE, based on prior experience before joining \companyname{}.} \\
\multicolumn{7}{l}{Senior Director* is a higher rank than a Senior Researcher.}
\end{tabular}}
\label{tab:demo}
\vspace{-4mm}
\end{table}

\textbf{Participant Recruitment.} 
We interviewed 24 participants at \companyname{}, as shown in Table \ref{tab:demo}. Using our internal directory, we first recruited 12 SDEs from the \companyname{}. We selected SDEs with senior experience or equivalent (or higher) with experience collaborating with DEs. This decision ensures participants had enough experiences and perceptions to provide in-depth and nuanced insights \cite{polkinghorne2005language}. We stopped recruiting participants after 12 interviews, as we reached saturation \cite{hennink2022sample} after the 10th interview, with the final two interviews conducted to confirm that no new themes emerged.

We then recruited DEs through snowball sampling by asking our SDE interviewees to connect us with DEs they know (1) whose title is not SDE and (2) who do not have a professional software engineering background but are collaborating with SDEs in co-developing software. We followed the same rule, requiring a ``Senior'' title or higher.  Demographic details are provided in the supplementary materials \cite{supply}.

These interviews were held remotely or on-site, depending on the participants' preferences. Each interview lasted 40-60 minutes and was audio-recorded for transcription and analysis. Participants signed a consent form before the interview following the \companyname{} ethics board. Participation in the study was voluntary, and participants received no compensation.


\textbf{Interview Analysis.} 
We analyzed the interviews incrementally. After each interview, we transcribed the audio recordings using Microsoft Teams and conducted thematic analysis \cite{kaur2024challenges, braun2012thematic} beginning with open coding to identify emergent themes \cite{howard2016glaser}. We took multiple passes over the transcribed data, allowing the codes to emerge naturally.

During the analysis, two authors met weekly to discuss emerging codes, develop a preliminary codebook, continuously review and update the codes, and resolve disagreements through negotiated agreement \cite{garrison2006revisiting}. Later, four authors participated in organizing and interpreting the data using affinity diagrams \cite{britz2000improving} and card sorting \cite{spencer2009card}. We discussed the rationale for applying specific codes and reached an agreement during the negotiated agreement process, validating the robustness of our codebook. The final codebook includes themes such as the expectations and frictions of SDEs and DEs. Supplementary documentation \cite{supply} provides the complete code list.

\textbf{Validity of Qualitative Findings.}
To validate our qualitative findings, as recommended by \citet{corbin2014basics}, we recruited two experts (one DE and one SDE) with experience in \DISD. They confirmed that our findings aligned with their own experiences. Three authors conducted two sessions with both experts to ensure our interpretation of the interview data aligned with their experiences. In the first session, we presented our interview questions, summaries, initial codebook, and anonymized quote samples and collected feedback and suggestions. We then refined the codebook accordingly. We followed up with the experts in another session to confirm the revisions until no significant changes.

\subsection{Study 2: Confirmatory Survey}

\textbf{Survey design.} 
Our survey began with a consent form approved by the university’s Institutional Review Board (IRB), followed by three demographic questions: experience in \DISD, the participant’s primary role in their most recent \DISD, and their gender. Participants who indicated no experience in \DISD were directed to exit the survey.

We then present questions regarding \emph{expectations and frictions} from our qualitative analysis. We asked participants to indicate their agreement with statements about their expectations of the other group (Eight expectations held by SDEs for DEs, and six expectations held by DEs for SDEs). We then asked participants how frequently they had encountered each friction (11 for SDEs and 10 for DEs) using a 4-point Likert scale  (``Never'', ``Rarely'', ``Occasionally'', ``Often''). The Likert scale for both questions was inspired by the Linux and Apache surveys to assess how frequently participants experienced interpersonal challenges when contributing to OSS \cite{LF2021, feng2023state}.

The questions regarding the expectations and frictions were separated based on the participant’s role. Specifically, participants were shown the questions for DE, unless they selected that their role is an SDE. We then manually validated their open-ended response that they were indeed in a DE role. See supplementary for survey questions \cite{supply}.


We also asked participants to rate their agreement with the statement: \textit{``My most recent experience with CDSD in software development was satisfying.''} This question was inspired by the Linux and Apache surveys on contributors' experiences \cite{LF2021, feng2023state}.


\textbf{Participant Recruitment.} 
We recruited participants by contacting developers from 19 OSS communities \cite{supply} across various domains (n=284), as well as an health department at \companyname{} (n=9). The OSS communities included Galaxy Project (astronomy), Bullet Physics (physics), RStudio (data science), and QuTiP (quantum computing). We also identified contributors with public email addresses in these repositories’ commit histories. Following ethical guidelines for data collection, we then emailed invitations, clearly explaining the survey’s purpose, voluntary nature, risks of the study, and data protection measures, with a consent form. All participants provided informed consent, and responses were anonymized in compliance with the General Data Protection Regulation (GDPR) and our IRB approval. This approach is consistent with prior software engineering research \cite{feng2022case, feng2025multifaceted}.

\textbf{Pilot survey.} 
Before distributing the survey, we conducted a pilot with three software engineering researchers, solely for collecting feedback on the survey design. The feedback helped us refine the phrasing of questions to improve clarity. For example, pilot participants suggested we clarify the concept of \DISD before starting the survey. In response, we explained \DISD in our recruitment email.


\textbf{Survey Participants Information.} 
In total, we received 512 responses. After removing invalid entries, we obtained 294 valid responses. Among them, 132 were identified as SDEs, 102 as DEs, and 58 selected ``Other'' and wrote in their role. 

We manually reviewed ``Other'' reposes. Many chose this option because their job titles were not as SDEs, but as DEs they are closely involved in software development. For example \textit{``I'm a bioinformatician with a biology background who has been doing software development''} [P\_DE273] (Survey participant). We excluded one participant who described their role as ``leadership,'' as we could not determine whether they were an SDE or DE. This resulted in a final dataset of 132 SDEs and 161 DEs.

Of the 293 participants, 240 identified as men, 27 as women, and 16 as non-binary. Most (77\%) had five or more years of cross-disciplinary collaboration experience. DEs came from diverse domains including finance, biology, health, psychology, physics, and engineering.



\textbf{Survey Analysis.} 
We conducted descriptive statistics for frictions and expectations. We also employed the Mann–Whitney U test \cite{nachar2008mann} to compare similar frictions reported by SDEs and DEs and to understand which group experienced similar paired frictions more frequently. 

We manually reviewed the open-ended responses related to frictions and expectations. These responses were rarely answered, and when provided, they either reinforced existing findings or offered additional details.

\label{sec:method}

\section{Unpacking Embedded Collaboration Through an Activity-Theoretic Lens (RQ1)}

In the following subsections, we unpack the \DISD{} collaboration model through the AT components, and then understand the expectations between SDEs and DEs.

\subsection{The embedded collaboration: \DISD}
We define \DISD{} as a model of embedded collaborative software development in which individuals from different disciplinary backgrounds contribute complementary expertise through overlapping roles, shared ownership of software artifacts, and blurred disciplinary boundaries across the development lifecycle.

We first conceptualize \DISD{} with AT components. The \textbf{\emph{Subjects}} are software development teams composed of SDEs and DEs from diverse domains with evolving roles. \textit{``My roles and responsibilities are often undefined, never clearly outlined at the edges''} [SDE2].  The \textbf{\emph{Object}} refers to the software artifacts developed throughout the process, ranging from prototypes, code modifications, and production-ready solutions \cite{de2003using}.   \textit{“There’s not an explicit handoff between the experimental code”} [SDE1]. \textit{“We write software that generates software”} [DE7].



\textbf{\emph{Division of Labor}} blurs as roles overlap, with tasks assigned by context rather than titles. \textit{``We kind of divided it up based on area of expertise, but the work was still very dependent on each other''} [DE6]. \textit{``It's very dependent, very dependent.'' } [DE10]. \textit{``I’ve been helping [DEs] set up models, build data pipelines, run inference, develop tools for parameter search, and fix code failures, there’s really no clear line''} [SDE4]. \textit{``never at the edges never appear''} [SDE3]. Teams draw upon heterogeneous \textbf{\emph{tools}}, often combining professional software development tools, programming languages, OSS infrastructure, bespoke scripts, and domain-specific pipelines. \textit{``We rely on OSS tooling, GCC, CMake, GitLab, GitHub, and integration frameworks like GitHub Actions, along with cloud services''} [DE10]. \textit{``We use GitHub for version control, testing discipline varies''} [SDE3]. \textit{``We have CI/CD systems set up, doing integration tests automatically''} [DE7].

\textbf{\emph{Rules}} are often implicit or inconsistently enforced, shaped by disciplinary norms and local practices. In some teams, formal engineering processes are vague: \textit{``On the team level, we actually do not currently have any kind of practices, best practices, or defined processes for managing the code base or handling pull requests''} [SDE10].  Elsewhere,
\textit{``We have tests, have the CI/CD system set up, doing integration tests automatically...you can only merge into the main branch if all the tests pass''} [DE10]. We observed that flexibility and accumulated inconsistencies remain \textit{``I'm used to it [relaxing code standard] now, but I wouldn't say it's the best code base for someone to understand''} [SDE8].


\textbf{\emph{Community}} is not hierarchical or modular but interwoven \textit{``When you're in a canoe, everybody paddles''} [DE7]. Collaboration is usually built on mutual trust, and success often depends less on formal processes. \textit{``When there is trust within the team and true collaboration, conversations happen organically. It's up to team members to recognize when something has been repeated too many times and to decide to automate those tasks to improve efficiency''} [SDE8].

The desired \textbf{\emph{outcome}} in \DISD{} is a successful project built through trust-based collaboration, extending from algorithm design [SDE6, SDE10, DE2, DE3, DE4, DE10], to libraries and APIs [SDE1, SDE4, SDE11, DE1, DE4, DE9, DE10], to applications [SDE1, SDE3, SDE5, SDE10, SDE12, DE1, DE2, DE5, DE7, DE12] and hardware systems [SDE9, DE2] (See supplementary for details \cite{supply}). 



\subsection{Emerging Expectation in \DISD}

\begin{table*}[!htbp]
\caption{Expectations (SE1 to SE8) that SDEs have for DEs (left) and DE expectations (E1 to E6) that DEs hold for SDEs (right), along with their mapping to AT components. Survey response greater than 15\% are shown for clarity. Color represents levels of agreement, ranging from red (strongly disagree) to dark green (strongly agree).}
\centering
\resizebox{0.89\textwidth}{!}{
\begin{tabular}{llllllll}
\hline
\rowcolor[HTML]{EFEFEF} 
\textbf{ID}                 & \textbf{AT Component}                    & \textbf{SDE expect DE to}                                                               & \textbf{Survey Responses (N=132)}                               & \textbf{ID} & \textbf{AT Component} & \textbf{DE expect SDE to}                                                 & \textbf{Survey Responses (N=161)}      \\ \hline
SE1                         & Object                                    & Validate reliability                                                                    & \five{1.52}{8.33}{7.58}{52.27}{30.3}                            & E1          & Division of Labor      & \begin{tabular}[c]{@{}l@{}}Conduct software \\ testing\end{tabular}       & \five{0.0}{2.48}{8.07}{59.01}{30.43}   \\
\rowcolor[HTML]{EFEFEF} 
SE2                         & Object                                    & \begin{tabular}[c]{@{}l@{}}Clear requests for re-\\ quirements\end{tabular}             & \five{1.52}{11.36}{13.64}{44.7}{28.79}                          & E2          & Division of Labor      & \begin{tabular}[c]{@{}l@{}}Improve code \\ quality\end{tabular}           & \five{0.62}{1.86}{16.15}{40.37}{40.99} \\
SE3                         & Community                                 & \begin{tabular}[c]{@{}l@{}}Take responsibility \\ for correctness\end{tabular}          & \five{1.52}{13.64}{13.64}{43.18}{28.03}                         & E3          & Community              & \begin{tabular}[c]{@{}l@{}}Provide technical \\ support\end{tabular}      & \five{0.62}{4.97}{14.29}{46.58}{33.54} \\
\rowcolor[HTML]{EFEFEF} 
SE4                         & Rules                                     & \begin{tabular}[c]{@{}l@{}}Write appropriate \\ documentation\end{tabular}              & \five{2.27}{9.09}{19.7}{48.48}{20.45}                           & E4          & Rules                  & \begin{tabular}[c]{@{}l@{}}Maintain \\ documentation\end{tabular}         & \five{0.0}{3.73}{18.63}{50.93}{26.71}  \\
SE5                         & Rules                                     & \begin{tabular}[c]{@{}l@{}}Follow agreed-upon \\ practices\end{tabular}                 & \five{6.06}{19.7}{25.0}{34.09}{15.15}                           & E5          & Division of Labor      & \begin{tabular}[c]{@{}l@{}}Make software \\ production-ready\end{tabular} & \five{0.62}{7.45}{16.77}{47.83}{27.33} \\
\rowcolor[HTML]{EFEFEF} 
SE6                         & Tools                                     & \begin{tabular}[c]{@{}l@{}}Understand\\ infrastructure\end{tabular}                     & \five{7.58}{24.24}{28.03}{34.85}{5.3}                           & E6          & Community              & \begin{tabular}[c]{@{}l@{}}Learn domain \\ knowledge\end{tabular}         & \five{4.35}{24.22}{41.61}{23.6}{6.21}  \\
SE7                         & Community                                 & \begin{tabular}[c]{@{}l@{}}Manage technical \\ dependencies\end{tabular}                & \five{11.36}{29.55}{23.48}{30.3}{5.3}                           & \multicolumn{4}{l}{}                                                                                                                                      \\
\cellcolor[HTML]{EFEFEF}SE8 & \cellcolor[HTML]{EFEFEF}Division of Labor & \cellcolor[HTML]{EFEFEF}\begin{tabular}[c]{@{}l@{}}Conduct code \\ reviews\end{tabular} & \cellcolor[HTML]{EFEFEF}\five{28.79}{35.61}{18.94}{13.64}{3.03} & \multicolumn{4}{l}{\multirow{-2}{*}{}}                                                                                                                    \\ \hline
\end{tabular}}

\label{tab:expect}
\vspace{-5mm}
\end{table*}

While \DISD promotes shared ownership and joint progress, it does not inherently align expectations. When roles blur and division of labor becomes fluid, implicit rules often replace explicit agreements. Understanding what each group expects from the other (\emph{subject}) is essential. Here, we unpack how expectations are formed, held, and prioritized in \DISD{}.  

\textbf{Expectations of SDEs.} 
Table~\ref{tab:expect} (SE1-SE8) lists the eight expectations that SDE holds for DEs through qualitative analysis and survey validations (N=132). Each expectation is mapped to the relevant AT components and color-coded based on the level of agreement from survey responses.

\textbf{(SE1) Shared Commitment to Reliability. (\emph{Object})}
SDEs expect DEs to validate and ensure the reliability of their contributions, with 82.6\% of survey respondents in agreement (52.27\% agree, 30.3\% strongly agree).  A robust and mutually trusted shared \emph{object} is critical to project success when responsibilities and ownership are fluid. \textit{``When you claim it’s gonna work''} [SDE6] \textit{``validating reliability is key''} [SDE12].


\textbf{(SE2) Clear requests for requirements. (\emph{Object})}
SDEs expect DEs to provide clear requests and articulate the \emph{object} of development, with 73.49\% survey agreement. \textit{``I expect all parties to participate in requirements definition and software''} [P\_SDE83], \textit{``I love to see clear expectations, but I know that’s not possible because things are so ambiguous''} [SDE8]. \textit{``I expect DEs to tell me, basically let me know how I can help''} [SDE5]. DEs also acknowledge, \textit{``I think a clear spec is important. I’ve seen this a lot more in the past where I think [SDEs] have struggled without clear guidance''} [DE2].


\textbf{(SE3) Taking Ownership and Maintaining Accountability. (\emph{Community})}
\textit{``Responsiveness, in both directions''} [P\_SDE82]. In a tightly coupled \DISD \emph{community}, where contributions are interdependent, SDEs expect DEs to take ownership and remain accountable for their contributions (71.21\% survey agreement).  \textit{``I expect them to own it, I would not maintain code that I did not make''} [SDE4]. However, this expectation is far from straightforward in practice and often controversial for both sides. The blurred boundaries and embedded collaborations make it difficult to define what ownership entails. \textit{``maintainability, reproducibility, and managing dependencies matters... That’s ownership''} [SDE9].

\textbf{(SE4) Documentation. (\emph{Rules})}
\textit{``Documentation is the minimum''} [SDE6] that helps harness complementary expertise across disciplines. SDEs expect DEs to provide clear and detailed documentation to grasp domain-specific logic and identify where they can help and how to move forward. (68.93\% agreement). \textit{``every project [is supposed to] have a great document that outlines exactly what has been done and why. In reality, that rarely happens''} [SDE5].

\textbf{(SE5) Engineering Practices. (\emph{Rules})}
SDEs expect DEs to \textit{``follow more engineering practices''} [SDE6]; especially as participants mentioned \textit{``Now we have a large group where many people work on the code base''} [DE4]. However, SDE and DE are frustrated that \textit{``they're not engineers and don't care about engineering''} [SDE11]. \textit{``their bar and their standard for software engineering and best practices are very different from a DE’s bar and standards''} [DE1], 



\textbf{(SE6) Infrastructure Awareness. (\emph{Tools})}
SDEs expect DEs to understand how their work fits into existing infrastructure pipelines. \textit{``[DEs] don’t tend to use [Tool], even though it’s already widely adopted in many applications''} [SDE6]. \textit{``I think they expect me to understand their tools. You know, they love their tools''} [DE9]. When DEs are unfamiliar with or excluded from the toolchains, it can lead to miscommunication, integration delays, and mounting frustration. SDEs then need to spend additional time mentoring them. \textit{``Kubernetes does all of the orchestration of the resources. I don’t know how to use Kubernetes… I didn’t develop it, but I have a strong dependency on it''} [DE2]

\textbf{(SE7) Dependency Management. (\emph{Community})}
SDEs also expect DEs to manage dependencies of their contributions to prevent conflicts during co-development, especially for some projects with a large developer base \textit{``we share that code base with something like 40 or 50 people''} [DE3].  As two SDEs from different teams (one from the US, the other from the UK) succinctly put it: \textit{``don’t step on each other’s toes''} [SDE2, SDE3]. \textit{``if ten people are modifying the same code base unless you have all of it organized, it’s going to be chaotic''} [SDE5].

\textbf{(SE8) Code Review (\emph{Division of Labor})}
We observed a contested expectations for code review, even among SDEs. Whether DEs participate in code review often depends on the trust built during collaboration. On one hand,  \textit{``we code review everything, even the MATLAB things''} [DE7], on the other hand \textit{``We’re not going to have… somebody [other DEs] who doesn’t have the experience or capability to really... they might want to, but we might just be like, no, that’s not a good idea''} [DE8]. Our survey also revealed this divide, with 28.79\% of participants disagreeing with letting DEs participate in code review. Such contradiction often occurs because SDEs assume DEs would make trade-off decisions that unintentionally introduce technical debt. \textit{``Oh man, I don't even think we think about quality''} [SDE10]


\textbf{Expectations of DEs.} Table~\ref{tab:expect} (right side) lists the expectations that DEs hold for SDEs.


\textbf{(E1) Testing Responsibility. (\emph{Division of Labor})}
With 89.44\%) agreement, DEs expect that SDEs should take responsibility for conducting software testing for their contributions. \textit{``validation or testing is expected [from SDEs]''} [DE1]. \textit{``I would expect [SDEs] to put a lot of thought into testing practices''} [DE11]. In practice, however, SDEs often struggle in finding a fine line between upholding quality standards and being perceived as overbearing \textit{``[DEs] don’t want us to policing them and make their life hard''} [SDE5].

\textbf{(E2) Improving Code Qualities. (\emph{Division of Labor})}
DEs often expect SDEs to improve, refactor, and polish code, \textit{``I expect SDEs to speed up and professionalize my code''} [P\_SDE214]. \textit{``They are good at code''} [DE10]. SDEs disagree with such task distributions \textit{``The DEs will need to spend a lot of their time learning...how to be engineers''} [SDE3].



\textbf{(E3)  Infrastructure Support and Technical Stewardship. (\emph{Community})}
DEs also expect SDEs to provide not only infrastructure support and technical stewardship.  \textit{``I expect SDE make life as easy as possible for both domain experts and themselves''} [DE1].\textit{``We expect [SDEs] develop a much better solution than we could, and it will be far more maintainable''} [DE2]. Yet SDEs often navigate a contradiction, while they are expected to provide infrastructure and technical support, their guidance can easily be perceived as controlling or manipulating. \textit{``[SDEs] really like to introduce [infrastructures] that I have to use. I have no choice''}[DE10]. 


\textbf{(E4) Documentations. (\emph{Rules})}
Both groups recognize the importance of appropriate code documentation. \textit{``Clean communication, as there are lots of things only some of us know. Translating jargon between fields is also important''} [P\_DE96]. \textit{``Without documentation, I can't understand how these things fit together and what it does''} [DE4].

\textbf{(E5) Elevating, Productizing, Scaling Software. (\emph{Division of Labor})}
DEs have expectations on productizing, scaling, and improving software. \textit{``I want it to be bug-free. I want it to be the latest version of everything... so my software doesn’t get outdated pretty soon''} [DE12]. However, SDEs often resist this assumed division of labor, \textit{``No, I will not impose how they write their code, so I cannot have that responsibility''} [SDE4].

\textbf{(E6) Learning and Mutual Adaptation. (\emph{Community})}
As SDEs expect DEs to invest time learning and practicing software engineering skills, DEs hold similar expectations in reverse, where each group presumes the other will take the initiative in bridging the gap. \textit{``[I] expect software engineers to learn domain knowledge and understand the [domain] aspect''} [DE5]. \textit{``So, for example, [SDE] in my team didn't know anything about medical imaging when they started. But he knows a lot now because he's had to''} [DE10].



\label{sec:RQ1}

\section{Frictions in \DISD (RQ2).}

When expectations are asymmetrical or unspoken, they give rise to frictions in collaboration \cite{he2021asymmetries, bozeman2016research}. These frictions typically manifest from (1) unmet expectations, where developers fail to fulfill assumed \emph{roles}, \emph{rules}, \emph{objects}, or \emph{division of labor} \cite{castaner2020collaboration, weiss2018objective}, or (2) controversial expectations, where conflicting assumptions about \emph{rules} or \emph{division of labor} \cite{he2021asymmetries, mukherjee2015coopting}. 

This section unpacks how these frictions surface across \DISD{} teams. We identified 21 frictions reported by SDEs (11) and DEs (10). These frictions are categorized into: \emph{Misaligned Priorities and Adaptation Pressure}, \emph{Unclear Boundaries and Practices}, \emph{Missing Context}, and \emph{High Workload}.

As shown in Table~\ref{tab:challenges}, we mapped the expectations to frictions faced by SDEs and DEs. The Frequency columns show the survey responses (N=132 SDEs, N=161 DEs), where participants indicated how often they experienced each friction on a 4-point Likert scale (``Never, Rarely, Occasionally, Frequently''). 
Next we analyzed how frictions manifest differently between SDEs and DEs through a Mann–Whitney test. The $U$ statistics and effect size $(r)$ (Table \ref{tab:challenges}) show these collaboration imbalances, for instance, how one group’s infrastructural decisions may be experienced as cognitive or technical overload by the other (row 4, Table \ref{tab:challenges}).



\subsection{Emerging Frictions in \DISD}

\begin{table*}[!htbp]
\caption{The table maps frictions faced by SDEs and DEs across Activity Theory (AT) components. Frequencies are based on survey responses (N=132 SDEs, N=161 DEs), where participants indicated how often they experienced each friction, ranging from never to often (color-coded from yellow to red).  Only survey response values greater than 15\% are shown for clarity.}
\centering
\resizebox{0.9\textwidth}{!}{
\begin{tabular}{llllllll}
\hline
\rowcolor[HTML]{EFEFEF} 
\multicolumn{3}{c}{\cellcolor[HTML]{EFEFEF}\textbf{Software Developer}}                                                                                                                                                                                                                       & \multicolumn{3}{c}{\cellcolor[HTML]{EFEFEF}\textbf{Domain Experts}}                                                                                                                                                                                                                    & \multicolumn{2}{c}{\cellcolor[HTML]{EFEFEF}\textbf{U statistics}} \\ \hline
\rowcolor[HTML]{EFEFEF} 
\multicolumn{1}{c}{\cellcolor[HTML]{EFEFEF}\textbf{Expectation}}                       & \multicolumn{1}{c}{\cellcolor[HTML]{EFEFEF}\textbf{Frictions}}                                                             & \multicolumn{1}{c}{\cellcolor[HTML]{EFEFEF}\textbf{Frequency (N=132)}} & \textbf{Expectation}                                                                   & \multicolumn{1}{c}{\cellcolor[HTML]{EFEFEF}\textbf{Frictions}}                                                      & \multicolumn{1}{c}{\cellcolor[HTML]{EFEFEF}\textbf{Frequency (N=161)}} & \textbf{U statistics}                & \textbf{$r$}               \\ \hline
\multicolumn{8}{c}{\textbf{Struggling with Misaligned Priorities and Adaptation Pressure}}                                                                                                                                                                                                                                                                                                                                                                                                                                                                                                                                                                 \\
\rowcolor[HTML]{EFEFEF} 
\begin{tabular}[c]{@{}l@{}}SE1-SE8\\ (aggregated)\\ E1-SE6\\ (aggregated)\end{tabular} & \begin{tabular}[c]{@{}l@{}}SDE facing \\ conflicting \\ piorities \\ with DE\end{tabular}                                   & \four{6.82}{31.82}{45.45}{15.91}                                       & \begin{tabular}[c]{@{}l@{}}SE1-SE8\\ (aggregated)\\ E1-SE6\\ (aggregated)\end{tabular} & \begin{tabular}[c]{@{}l@{}}DE facing \\ conflicting \\ piorities \\ with SDE\end{tabular}                            & \four{9.32}{37.89}{39.13}{13.66}                                       & 11578.5 ns                           & 0.08                       \\
\begin{tabular}[c]{@{}l@{}}SE3, SE5, \\ SE7, E2, \\ E3, E5\end{tabular}                & \begin{tabular}[c]{@{}l@{}}SDE struggling with \\ technical debt due to  \\ DE's rapid iteration\end{tabular}               & \four{11.36}{24.24}{33.33}{31.06}                                      & \begin{tabular}[c]{@{}l@{}}E1, SE1, \\ SE3\end{tabular}                                & \begin{tabular}[c]{@{}l@{}}DE struggling with \\ SDE's testing \\ demands\end{tabular}                               & \four{24.22}{40.37}{24.84}{10.56}                                      & 14289.0 ***                          & 0.30                       \\
\rowcolor[HTML]{EFEFEF} 
\begin{tabular}[c]{@{}l@{}}SE5, E2, \\ E5\end{tabular}                                 & \begin{tabular}[c]{@{}l@{}}SDE facing resistance \\ from DE in adopting \\ Understanding SDE \\ practices\end{tabular}      & \four{18.94}{24.24}{34.09}{22.73}                                      & \begin{tabular}[c]{@{}l@{}}E2, SE5, \\ SE7\end{tabular}                                & \begin{tabular}[c]{@{}l@{}}DE finding it \\ challenging and\\ time-comsuming to \\ follow SDE practices\end{tabular} & \four{18.63}{31.06}{36.02}{14.29}                                      & 11525.0 ns                           & 0.07                       \\
SE6, E6                                                                                & \begin{tabular}[c]{@{}l@{}}SDE needing to adopt \\ new technologies and \\ languages for \\ collaboration\end{tabular}      & \four{9.09}{32.58}{38.64}{19.70}                                       & E3, SE6                                                                                & \begin{tabular}[c]{@{}l@{}}DE struggling with \\ numerous complex \\ SDE infrastructures\end{tabular}                & \four{22.36}{35.40}{29.19}{13.04}                                      & 12844.0 **                           & 0.18                       \\
\rowcolor[HTML]{EFEFEF} 
\multicolumn{8}{c}{\cellcolor[HTML]{EFEFEF}\textbf{Struggling with Unclear Boundaries and Practices}}                                                                                                                                                                                                                                                                                                                                                                                                                                                                                                                                                      \\
\begin{tabular}[c]{@{}l@{}}SE3, SE5\\ SE7, E1, \\ E2\end{tabular}                      & \begin{tabular}[c]{@{}l@{}}SDE struggling with \\ relaxed code quality \\ standards\end{tabular}                            & \four{21.21}{27.27}{30.30}{21.21}                                      & \begin{tabular}[c]{@{}l@{}}E2, E3, \\ SE3, SE5\\ SE7\end{tabular}                      & \begin{tabular}[c]{@{}l@{}}DE struggling in \\ adopting SDE's high \\ code standards\end{tabular}                    & \four{33.54}{42.24}{21.74}{2.48}                                       & 13956.0 ***                          & 0.27                       \\
\rowcolor[HTML]{EFEFEF} 
\begin{tabular}[c]{@{}l@{}}SE3, SE7, \\ SE8, E5\end{tabular}                           & \begin{tabular}[c]{@{}l@{}}SDE Struggling with \\ Blurred Ownership \\ and Limited Authority\end{tabular}                   & \four{37.12}{41.67}{15.15}{6.06}                                       & \begin{tabular}[c]{@{}l@{}}E5, SE3, \\ SE7, SE8\end{tabular}                           & \begin{tabular}[c]{@{}l@{}}DE Struggling with \\ Blurred Ownership \\ and Limited Authority\end{tabular}             & \four{68.94}{12.42}{11.80}{6.83}                                       & 13428.5 ***                          & 0.23                       \\
\begin{tabular}[c]{@{}l@{}}SE1, SE2, \\ E3\end{tabular}                                & \begin{tabular}[c]{@{}l@{}}SDE receiving \\ ambiguous and \\ evolving requirements\end{tabular}                             & \four{3.79}{25.76}{39.39}{31.06}                                       &                                                                                        &                                                                                                                      &                                                                        &                                      &                            \\
\rowcolor[HTML]{EFEFEF} 
\multicolumn{8}{c}{\cellcolor[HTML]{EFEFEF}\textbf{Struggling with Missing Context}}                                                                                                                                                                                                                                                                                                                                                                                                                                                                                                                                                                       \\
\begin{tabular}[c]{@{}l@{}}SE4, E4,\\ E6\end{tabular}                                  & \begin{tabular}[c]{@{}l@{}}SDE lacking sufficient \\ domain knowledge \\ resource\end{tabular}                              & \four{17.42}{42.42}{35.61}{4.55}                                       & \begin{tabular}[c]{@{}l@{}}E4, E6,\\ SE4\end{tabular}                                  & \begin{tabular}[c]{@{}l@{}}DE struggling due to \\ SDE's insufficient \\ domain knowledge\end{tabular}               & \four{29.81}{33.54}{28.57}{8.07}                                       & 11554.0 ns                           & 0.08                       \\
\rowcolor[HTML]{EFEFEF} 
\begin{tabular}[c]{@{}l@{}}SE5, SE6, \\ E3, E5\end{tabular}                            & \begin{tabular}[c]{@{}l@{}}SDE struggling due to \\ DE's lack of knowledge\\ and proficiency \\ in using tools\end{tabular} & \four{17.42}{40.91}{38.64}{3.03}                                       & \begin{tabular}[c]{@{}l@{}}E3, E5\\ SE5, SE6\end{tabular}                              & \begin{tabular}[c]{@{}l@{}}DE struggling in \\ adopting SDE \\ development tools \\ (e.g., IDE)\end{tabular}         & \four{52.17}{34.16}{11.80}{1.86}                                       & 15198.5 ***                          & 0.37                       \\
SE4, E4                                                                                & \begin{tabular}[c]{@{}l@{}}SDE struggling to \\ understand the \\ codebase due to DE's \\ poor documentation\end{tabular}   & \four{11.36}{35.61}{34.09}{18.94}                                      & E4, SE4                                                                                & \begin{tabular}[c]{@{}l@{}}DE struggling to \\ understand SDE \\ code due to poor \\ documentation gaps\end{tabular} & \four{19.88}{34.78}{37.27}{8.07}                                       & 12232.5 *                            & 0.13                       \\
\rowcolor[HTML]{EFEFEF} 
\multicolumn{8}{c}{\cellcolor[HTML]{EFEFEF}\textbf{Struggling with High Workload}}                                                                                                                                                                                                                                                                                                                                                                                                                                                                                                                                                                         \\
\begin{tabular}[c]{@{}l@{}}SE1-SE8\\ (aggregated)\\ E1-SE6\\ (aggregated)\end{tabular} & \begin{tabular}[c]{@{}l@{}}SDE overwhelmed \\ with high workloads \\ and development \\ demands\end{tabular}                & \four{8.33}{40.15}{34.09}{17.42}                                       & \begin{tabular}[c]{@{}l@{}}SE1-SE8\\ (aggregated)\\ E1-SE6\\ (aggregated)\end{tabular} & \begin{tabular}[c]{@{}l@{}}DE struggling to \\ manage both domain \\ tasks and SDE tasks\end{tabular}                & \four{21.12}{29.81}{37.27}{11.80}                                      & 11785.5 ns                           & 0.09                       \\
\rowcolor[HTML]{EFEFEF} 
\multicolumn{8}{l}{\cellcolor[HTML]{EFEFEF}\begin{tabular}[c]{@{}l@{}}Significance markers: \textit{ns} = not significant; * = \textit{p} $<$ 0.05; ** = \textit{p} $<$ 0.01; *** = \textit{p} $<$ 0.001; **** = \textit{p} $<$ 0.0001.\\ Effect size thresholds ($r$): small = 0.1, medium = 0.3, large = 0.5.\end{tabular}}                                                                                                                                                                                                                                                                                                                              \\ \hline
\end{tabular}}
\label{tab:challenges}
\vspace{-5mm}
\end{table*}

\textbf{Frictions from Conflicting Priorities.} 
The different backgrounds and expertise are the root causes of friction. \textit{``Different philosophical views... in terms of how to approach software development between [SDEs] and [DEs]... ''} [DE1]. These frictions reflect not just single misunderstandings but aggregated misalignment in expectations about what software development should prioritize. For example, \textit{``All they know is that they need to be able to read data, process it, produce some output, and then do it as many times as possible until they get the correct result, [DEs] are not focused on what platforms they should be using or how they should be doing this most efficiently''} [SDE5]. On the other side, \textit{``They wanted something very easy to extend from what they've already done... but you lose a lot of [domain] essential''} [DE5].

\textbf{Frictions from Technical Debt.}
An SDE reported struggling with technical debt due to rapid iteration, \textit{``The real cost of building as rapidly as possible is that we left a lot of technical debt. But when the goal is to iterate and prove whether something can be done quickly, you don't really have a choice''} [SDE7]. From a DE's perspective, \textit{``I just wanna like test out ideas quickly, When I push my commits and want to see my changes''} [DE9]. Our U test on survey responses indicated a statistically significant difference where SDEs reported struggling with technical debt more often than DEs reported struggling with SDEs’ testing demands, with a moderate effect size ($U = 14{,}289$, $p < 0.001$, $r = 0.30$). \textit{``DEs choosing technical shortcuts knowing that the ultimate labor/time cost of those choices would fall on SEs...''} [P\_SDE128].

\boldification{SE practices is another tension reported by SE}
\textbf{Frictions from Enforcing (or Resisting) SE Practices.} 
Unsurprisingly, SDEs and DEs diverge sharply in adopting software engineering practices  \textit{``[DEs] write code that’s too brittle because they work fast and just want to get to the end result. [SDEs] write code that’s too rigid because they care about the code being very clean.''} [DE1]. \textit{``Doing software engineering properly is a hard job, and you can't ask someone who's already overwhelmed.''} [SDE3].  Even DEs expressed frustration with peers who neglect agreed-on practices \textit{``As a DE and also lead programmer, I frequently experience frustration with [DE] colleagues who do not comply with agreed upon best practices for software development''} [P\_DE265].


\textbf{Frictions from Adapting Unfamiliar Tools/Infrastructures.} 
Both SDEs and DEs mentioned struggling to adopt new languages or infrastructures they are unfamiliar. \textit{``I have no choice because [the tool] is part of the infrastructure... it can create a lot of work''} [SDE9]. Our survey analysis revealed that such struggle is reported significantly more often by SDEs ($u =$ 12,844, $p <$ 0.01, $r =$ 0.18). \textit{``When you join a new team, you would adapt to a new technology process and language altogether''} [SDE5]. \textit{``I got one language that is maybe worse than another one''} [SDE6]. DEs mentioned infrastructures that they do not understand take a lot of time to learn \textit{``The system we use is quite sophisticated, but by design, a lot of the dependencies we rely on are hidden away from us''}[DE1].

\textbf{Frictions from Code Quality.} 
Code quality is another persistent source of friction, where SDEs think DEs lag in maintaining code quality, while DEs find writing high-quality code challenging. \textit{``I think this is gonna create constant struggle in our code base, I think their bar and their standard for software engineering and best practices are very different than our bar and standards''}[DE1].  While SDEs often push for modularized and maintainable code that can be easily adapted or reused, DEs often see them as an unnecessary time-wasting effort. \textit{``Sometimes I find it unnecessary to follow the required structure. I wish they could find a little bit of a sweet spot''}[DE9]. Our survey results revealed that SDEs report encountering code quality issues significantly more often than DEs struggling to adopt SDEs’ code standards, with a nearly moderate effect size ($u =$ 13,956, $p <$ 0.001, $r =$ 0.27).



\boldification{Both groups struggle with authority over the code base}
\textbf{Frictions from Blurred Ownership.} 
Both SDEs and DEs expressed frustration over blurred ownership. \textit{``Because the repository right now is controlled by the [DEs], not by us [SDEs], there are no strong engineering practices around how the model is maintained and updated''} [SDE7]. \textit{``Everyone owns a different chunk of code''} [DE3]. \textit{``it’s not my responsibility''} [DE7]. \textit{``Even though [those are] smaller pieces, they go further and they are not necessarily owned by me [anymore]''} [SDE6].


\textbf{Frictions from Ambiguous Requirements.} 
SDEs often struggle with ambiguous and evolving requirements. over 70\% of SDE participants reported often/occasionally facing this friction. \textit{``The project objectives at the start usually do not match the project objectives at the end''}[P\_SDE114] \textit{``Occasionally DEs did not fully understand the proposed solutions before beginning development''}[P\_SDE189]. Ultimately, these ambiguous requests weaken trust and negatively impact collaborations \textit{``When the code you've just written could be scrapped, and you have to rewrite it in a couple of weeks because they decide: `that's not exactly what we want; we want it to work a little differently''} [SDE10].

\textbf{Frictions from Knowledge Silos.} 
Both DEs and SDEs acknowledge that insufficient domain knowledge is a friction. \textit{``they will never understand the model, they just understand what the output is''} [DE10]. \textit{``I am receiving [domain knowledge] from the team I’m working on, but tiny bits of information. They just don’t want to share the whole idea of the project''} [SDE12]. \textit{``I think we ask [SDE] to Google it, there’s not like a textbook or something''} [DE5].  Similarly to software engineering knowledge. Our survey responses also revealed that SDEs more frequently struggle due to DEs’ lack of knowledge and proficiency in using software development tools, with a moderate effect size ($u =$ 15,198, $p <$ 0.001, $r =$ 0.37). \textit{``These tools have a much steeper learning curve and very little interoperability''} [P\_DE272].

\textbf{Frictions from Documentation Debt.} 
\textit{``I laugh when you say documentation''} [SDE7]. While the knowledge gaps could potentially be mitigated through up-to-date documentation \cite{parnas2010precise}, the reality is that poor documentation often exacerbates the problem of development \cite{zhi2015cost}. \textit{``An enormous part in our codebase, I don't know what exactly it means and what it's doing because I'm not a software engineer''} [DE1]. Interestingly, our survey results revealed that SDEs encounter documentation-related frictions slightly more often than DEs, with a small effect size ($u =$ 12232.5, $p <$ 0.05, $r =$ 0.13). \textit{``Part of an interdisciplinary project is improving the documentation so more developers with less domain (and software) expertise and more users with less domain (and software) expertise can use it''}[P\_SDE15].

\textbf{Overwhelmed Developers.} 
These frictions are further exacerbated by the overload experienced by both SDEs and DEs. \textit{``We need to maintain both [DEs] and product work at the same time''}[DE9]. \textit{``We encounter issues about workloads''}[SDE4].  \textit{``I have to take my free time} [P\_SDE199].

%

\subsection{Prevalence and Impact of Friction}

Our survey revealed that frictions in \DISD{} are widespread. Recall we also asked participants to rate their agreement with satisfaction statements regarding their recent \DISD project; 18.94\%  of SDEs and 15.53\% of DEs reported being not satisfied. Our U-tests on paired friction revealed that SDEs consistently reported experiencing identified frictions more frequently than DEs for all statistically significant cases. This trend may explain why SDEs reported lower satisfaction levels compared to DEs. \textit{``I'm the least liked person on the team''} [P\_SDE84]. \textit{I was told I'm not a team player''} [SDE2].

We also list here the most common friction reported by SDEs and DEs. Among SDEs, eight out of 11 frictions were reported by over 50\% of respondents as occurring at least occasionally. The top three frictions were:  
(1) ambiguous and evolving requirements (31.06\% often, 39.39\% occasionally);
(2) technical debt from rapid iteration (31.06\% often, 33.33\% occasionally);   
(3) \textit{conflicting priorities} (15.91\% often, 45.45\% occasionally). Two frictions were reported by more than 50\% of DE respondents: (1) \textit{conflicting priorities with SDE} (13.66\% often, 39.13\% occasionally), (2) \textit{struggling to follow SDE practices} (14.29\% often, 36.02\% occasionally).

\label{sec:RQ2}

\section{Discussion and Remarks}

\begin{figure}[!htbp]
    \centering
    \includegraphics[width=0.8\columnwidth]{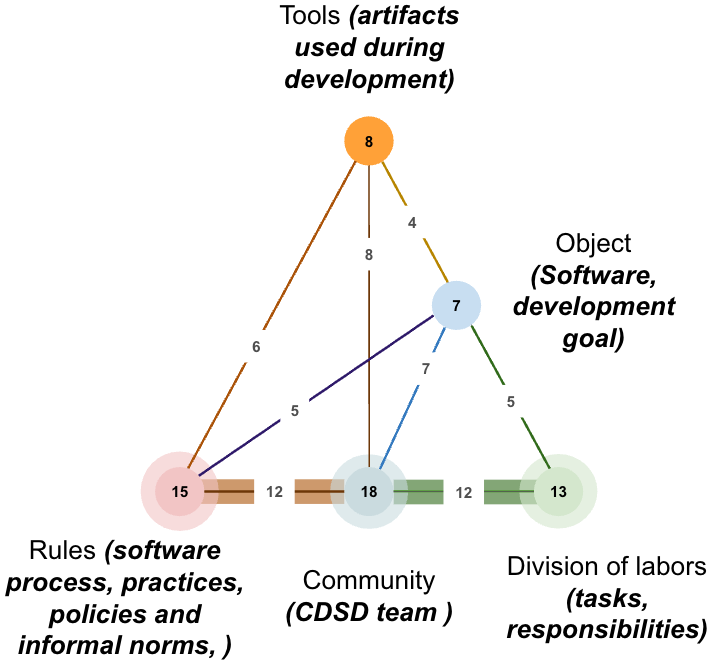}
    \caption{AT component co-occurrence network}
    \label{fig:atcluster}
    \vspace{-4mm}
\end{figure}

Through our explorations in \DISD, we learned about developers'--DEs and SDEs--experiences and their frustrations. This section discusses friction hotspots that can inform future research, infrastructure design, and practical guidance for practitioners. We constructed a network visualization based on relationships among AT components as shown in Figure~\ref{fig:atcluster}.

Each vertex in the network represents an AT component, as introduced in Section IV-A. The weight of a vertex corresponds to the number of frictions (Table~\ref{tab:challenges}) that involve at least one expectation mapped to that component via the expectation-to-component mapping (defined in Table~\ref{tab:expect}). Specifically, let $C$ be the set of all frictions, and $f(e)$ is the AT component for expectation $e$. Then $w(v) = |\{ c \in C \mid \exists e \in c \text{ s.t. } f(e) = v \}|$. For example, among 21 frictions, expectations related to \textit{Community} occurred 18 times (node weight).  Each edge represents the co-occurrence of two AT components within the same friction. For example, the edge between \emph{Rules} and \emph{Community} weights 12, indicating that these two components co-occurred in 12 frictions. Specifically, if $f(c)$ denotes the set of AT components associated with friction $c$, then the weight of an edge between $v_i$ and $v_j$ is: $w(v_i, v_j) = |\{ c \in C \mid v_i \in f(c) \text{ and } v_j \in f(c) \}|$.

We exclude the \emph{Subject} and \emph{Outcome} components since \emph{Subject} consistently represents the developers (SDE or DE) and appears in all frictions, and \emph{Outcome} is the shared goal of successful software development. These two components do not offer discriminative insight into friction patterns.

\textbf{Implications for Future Research.}
%
As shown in Figure \ref{fig:atcluster}, \emph{Community} is heavily implicated with strong paths to \emph{Rules} and \emph{Division of Labor}. Future research could: (1) Reimagine hybrid collaboration models that integrate AT with other frameworks such as Distributed Cognition (DC) \cite{hutchins2000distributed}, which aims to understand how thinking and knowledge are distributed across people, environments, and tools. Such integration could inform  decision-revision processes or adaptive testing workflows that accommodate rapid experimentation while preserving engineering integrity. For example, in CDSD, shared artifacts like decision logs or experiment dashboards could be studied both as mediating tools in AT (surfacing contradictions) and as cognitive scaffolds in DC (distributing decision-making across SDEs and DEs). (2) Investigate how mutual onboarding practices can be designed and evaluated through the lens of Communities of Practice \cite{wenger2011communities}, which views learning as a social process rooted in shared practice, to enhance knowledge sharing and accelerate productivity.


\textbf{Implications for Infrastructure Designers.}
Our study also reveals a gap between technical infrastructure and socio-technical realities in \DISD. Tools that merely enforce quality without scaffolding collaboration exacerbate the divide. We suggest:(1) Automated annotation systems that allow DEs to explain the rationale behind code and enable SDEs to annotate code with explanations focused on maintainability. (2) Embedded friction indicators integrated into IDEs or CI/CD pipelines that can automatically flag issues when expected practices are missing, which can, in turn, help teams identify and resolve frictions early.
(3) Finally, Large Language Model (LLM) mentors \cite{xu2022systematic, feng2024guiding} can serve as a knowledge bridge if LLMs can adapt their recommendations based on the developer’s background and the required artifacts reducing frictions clarification of expectations around rules and tools.

\textbf{Implications for Practitioners.}
We observed that the successful \DISD teams rely on implicit trust, implicit mentoring \cite{feng2022case}, and continuous renegotiations of roles with equal voice. We provide a roadmap for identifying which components are most prone to friction, helping \DISD teams prioritize where to invest in alignment, communication, and scaffolding before a project begins, rather than leaving collaboration norms to emerge or relying on one side to compromise for the other. We suggest that practitioners and governance teams consider the following actions: (1) Implement expectations alignment before initiating the project where teams should explicitly state what each group expects from the other using our AT framework, especially rules and division of labor, and establish mechanisms to revisit and renegotiate expectations; (2) Conduct periodic co-ownership check-ins to realign evolving roles, clarify responsibilities, and reaffirm mutual goals.

\textbf{Threats to validity.}
%
In a mixed-method empirical study in a heterogeneous domain like SE, theory plays a critical role in guiding analysis and avoiding purely descriptive work \cite{stol2015theory}.
We leveraged AT as our theoretical lens as it enables systematic analysis of the human–tool–community interactions and has been successfully applied in SE research \cite{de2003using}.

We acknowledge that the interviews were conducted within \companyname{}, yet the participants are from diverse fields, organizations, projects, and regions, with a balanced representation of SDEs and DEs.  We also conducted two strategies to mitigate the effects of research bias, confirmation bias, and interpretive subjectivity \cite{onwuegbuzie2007validity}. First, we invited external researchers to collaborate on the qualitative analysis. 
Second, we employed a mixed-methods design to enhance the credibility, transferability, and resonance of our findings \cite{storey2024guiding, tracy2010qualitative}. We conducted a large-scale survey beyond \companyname{} to validate our qualitative constructs across diverse organizational and community settings.

Conceptual clarity is essential for grounding theory-oriented \cite{stol2015theory}, so we used AT as the theoretical lens for our data collection and analysis efforts. While AT provided a structure for understanding collaboration frictions, it may have constrained our focus, as some frictions could fall outside the scope of what AT captures. However, the validation survey with developers beyond our interview sample revealed no additional friction categories, suggesting our theoretical framing was reasonably comprehensive.


Regarding the qualitative analysis, no fixed number of interviews is required; the focus is on achieving data saturation \cite{sebele2020saturation, saunders2018saturation, caine2016local}. While the required sample size remains debated \cite{townsend2013saturation}, prior empirical work shows that many themes often emerge within the first six interviews \cite{guest2006many}. Qualitative methodology guidance also notes that some approaches, such as phenomenology, commonly employ small (3-10) samples \cite{creswell2016qualitative}. Concerning different perspectives, our study achieved saturation.

\textbf{Conclusion and Future Work.}
In this work, we used AT as an analytical lens to investigate the dynamics of \DISD. By learning from both SDEs and DEs, we identified eight expectations that SDEs hold for DEs and six expectations that DEs hold for SDEs. When expectations are unmet or contested, frictions arise. We identified 21 frequent frictions and understand how they manifest in practice.

As \DISD{} continues to emerge, addressing the frictions identified in this study will be critical to improving effectiveness and efficiency for software development. Our future work plans to design LLM-powered tools to help shape more adaptive, inclusive, and effective \DISD{} environments.

\section{Acknowledgment}
We thank all interview and survey participants for their time and insights. This work was partially supported by NSF Grant No. 2303043.

\label{sec:discussion}



\begingroup
\scriptsize
\bibliographystyle{IEEEtranN}
\bibliography{bib.bib}

\end{document}